\documentstyle[epsf,epsfig]{aipproc}
\begin{document}
{\hfill LBNL-40495}
\vskip .01 in
{\hfill June, 1997}
\vskip -.1 in
\title{Coherent Photons and Pomerons in Heavy Ion Collisions}
\author{Spencer Klein$^1$ and Evan Scannapieco$^{1,2}$}
\address{$^1$Lawrence Berkeley National Laboratory, Berkeley, CA, 94720, USA \\
$^2$ Department of Physics, University of California, Berkeley CA 94720-7304, USA.}
\maketitle
\def\gamgam{$\gamma\gamma\,$}
\def\gampom{$\gamma P$}
\def\pompom{$PP$}
\vskip -.1 in
\begin{abstract}
Ultrarelativistic heavy ion beams carry large electromagnetic and
strong absorptive fields, allowing exploration of a variety of
physics. \gamgam, \gampom, and \pompom\ interactions can probe a huge
variety of couplings and final states.  RHIC will be the first heavy
ion accelerator energetic enough to produce hadronic final states via
coherent couplings.  Virtual photons from the nuclear EM fields can
interact in \gamgam\ interactions, which can be exploited to study
many particle spectroscopy and QCD topics.  Because the photon flux
scales as $Z^2$, \gamgam\ luminosities are large up to an energy of
about $\gamma\hbar c/R\approx 3$ GeV/c.  Photon-Pomeron interactions
are sensitive to how different vector mesons, including the $J/\psi$,
interact with nuclear matter.  $PP$ collisions rates are sensitive to
the range of the Pomeron.  Signals can be separated from backgrounds
by using cuts on final state isolation (rapidity gaps) and $p_\perp$.
We present Monte Carlo studies of different backgrounds, showing that
representative signals can be extracted with good rates and signal to
noise ratios.
\end{abstract}
\vskip -.14 in
\centerline{(Presented at the 6th Conference on the Intersections of Particle}
\centerline{and Nuclear Physics, May 27-June 2, 1997, Big Sky, Montana)}
\vskip .18 in
\section{Physics Processes}

When it is completed in 1999, the Relativistic Heavy Ion
Collider\cite{RHICLUM} (RHIC) will be energetic enough to produce
hadronic final states via \gamgam, \gampom, and \pompom\ interactions
that coherently couple to the nuclei as a whole.  The electromagnetic
field of the heavy nuclei can be considered as a flux, proportional to
$Z^2$ of almost-real Weisz\"acker-Williams virtual photons.  The
requirement that the photons couple to the entire nucleus limits the
maximum photon energy to $\gamma\hbar c/R$, about 3 GeV/c for gold
beams at RHIC. Thus, the maximum \gamgam\ energy is about 6 GeV; at
energies of a few GeV, the \gamgam luminosity will be comparable to
those of the next generation $e^+ e^-$ colliders.

\gamgam\ interactions can probe a wide variety of physics topics.
Particle coupling to two photons is a measure of their internal
charge; $q\overline q$ mesons couple strongly, but glueballs and mixed
states ($q\overline qg$) should have much smaller couplings.
Nonobservation in \gamgam\ collisions is therefore an important
criteria for identification of glueball candidates\cite{paar}.
\gamgam collisions can also probe spin 0 or 2 mesons with exotic
quantum numbers.  One interesting example is the near-threshold
$\rho^0\rho^0$ resonance, produced in \gamgam\ collisions at $e^+e^-$
colliders.  According to the particle data book, ``This process has
not been explained by models in which only conventional resonances
dominate''\cite{pdg}.

\gamgam\ collisions also produce large numbers of lepton and hadron
pairs.  Lepton pairs can be used to measure the \gamgam\ and relative
hadronic luminosity, and to search for nonlinear QED effects due to
the large coupling ($Z\alpha\approx 0.6$).  Meson pairs can be used to
measure form factors and for a variety of other QCD studies, and baryon
pair production can test diquark based models\cite{pairb}.

Several authors have calculated the \gamgam\ luminosity produced by
heavy ion colliders\cite{candj}\cite{baurandff}\cite{hencken}. The
\gamgam\ luminosity can be found by convoluting the photon fields of
the two nuclei, and integrating over impact parameter $b$ greater than
twice the nuclear radius $R_A$.  The latter criteria avoids events
where hadronic collisions overshadow the \gamgam\ interaction, but
cuts the usable luminosity by about 50\%.  Figure~1 compares the
usable \gamgam\ luminosities for RHIC Au, Cu, and I beams to those of
CLEO at CESR and LEP2.  The lighter nuclei benefit from the higher
$AA$ luminosity and beam energy, and smaller nuclear radius, which
more than compensate for the reduced $Z$. Because of the photon energy
cutoff, final states are produced quite centrally, as shown in Fig.~1.
Final states from $e^+e^-\rightarrow\gamma\gamma$ have a wider $y$
distribution, and so, for a given setup, the experimental acceptance
is lower.

\begin{figure} 
\centerline{
\epsfig{file=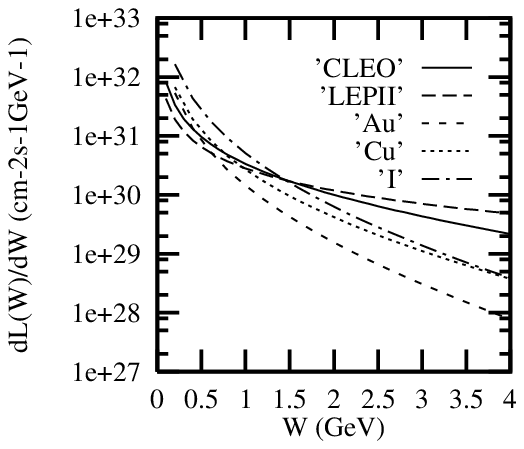,height=1.8in,width=2.6in} 
\epsfig{file=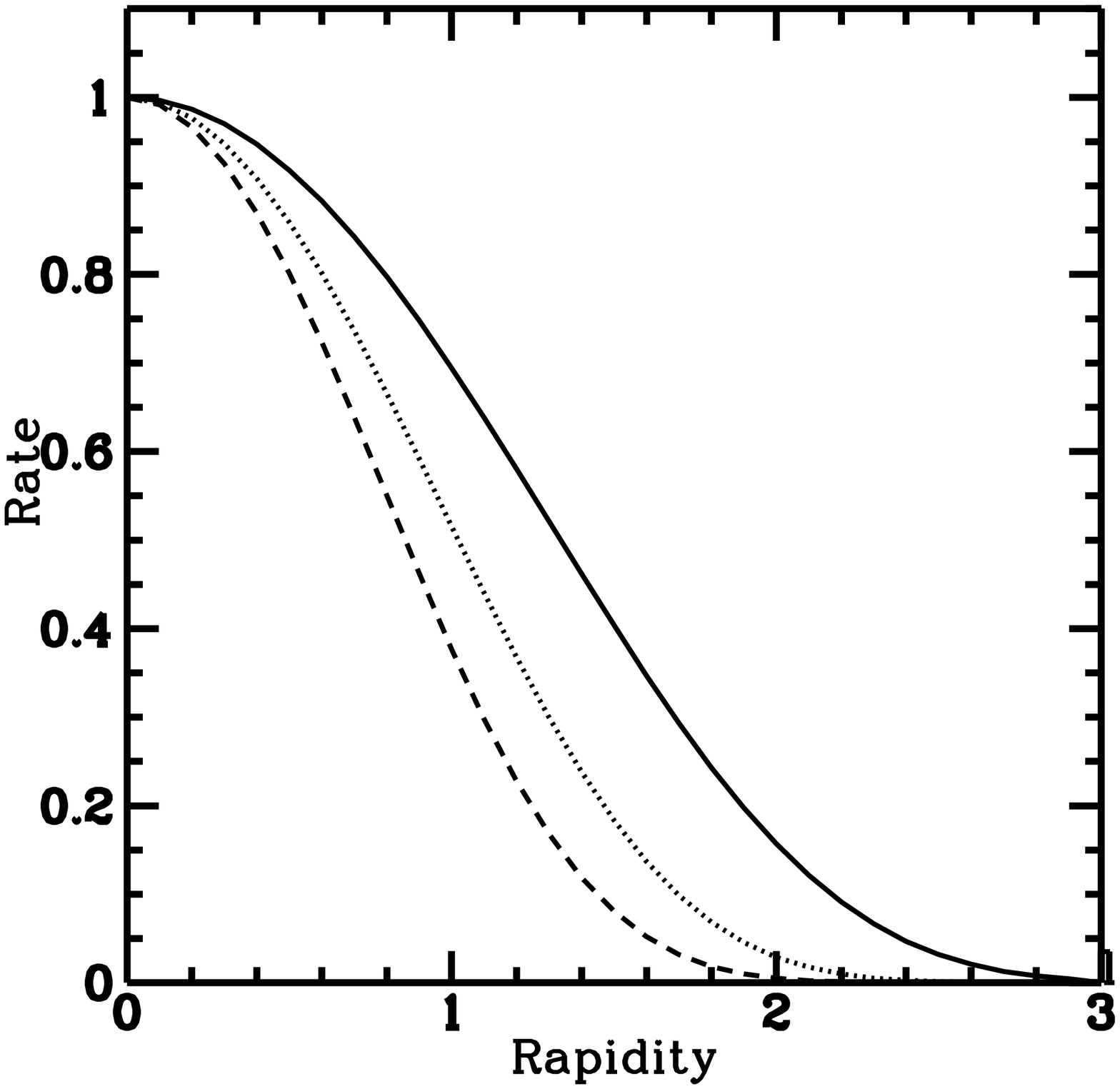,angle=0,height=1.9in,width=2.2in}
}
\caption{Left: Comparison of \gamgam\ luminosities at RHIC, for Au, I
and Cu beams,  with those of CESR (CLEO) and LEP II.  Right:
Rapidity distribution of produced particles for gold on gold collisions
with W = 1 (solid), 2 (dotted) and 3 (dashed) GeV.}
\label{fig:luminosity}
\end{figure}
Due to the nuclear form factor, the photons are almost real, with a
maximum $Q^2$ given by the nuclear size, about (30 MeV/c)$^2$ for
gold.  This cutoff limits the perpendicular momentum of the photons,
$p_\perp < \hbar c/R$; this helps separate coherent from incoherent
interactions.  This is illustrated in Fig 2.  The $Q^2$ limit will
slightly reduce the rate of pair production near
threshold\cite{baur2}.

In addition to \gamgam\ collisions, the virtual photons can also
couple to the Pomeron field of the other nucleus.  The Pomeron can be
thought of as representing the absorptive part of the nuclear
potential.  \gampom\ interactions using proton targets were studied
extensively at HERA.  RHIC can study these interactions in a nuclear
environment.  One reaction of interest is $\gamma P\rightarrow V$,
where $V$ is a vector meson.  In the Vector Dominance Model, the
photon can be considered to fluctuate into a spin 1 $q\overline q$ state,
which then interacts with the nucleus.  By studying production rates
of various sized mesons, and varying the nuclear radius by changing
the beams, the interactions between quark pairs and nuclei can be
studied\cite{brodsky}.  The kinematics are similar to \gamgam\
processes, so similar detection techniques can be used. RHIC will
reach higher energies and luminosities than earlier NMC\cite{NMC} and
Fermilab E-665\cite{E665} studies, producing 100,000's of exclusive
$\rho$ and $\phi$ mesons per year, large numbers of excited states,
and the $J/\psi$.  The latter is of special interest because the
large quark mass may require perturbative treatment.

In addition to photon physics, double Pomeron interactions can be
studied.  Unobscured \pompom\ interactions can only occur in the
impact parameter range $2R_A+2R_P > b > 2R_A$, where $R_P$ is the
range of the Pomeron.  The \pompom\ cross section is thus sensitive to
the range of the Pomeron.  As \gamgam\ and \pompom\ final states have
similar kinematics, a statistical separation is required for this
measurement.  The relative rates will depend on $A$; \pompom\
couplings will dominate for small nuclei and \gamgam\ for large.

In many cases, the same final state can be produced through more than
one intermediate state.  Therefore, the possibility of interference
exists.  One place where it should occur is between the reactions
$\gamma P \rightarrow V\rightarrow e^+e^-$ and
$\gamma\gamma\rightarrow e^+e^-$\cite{leith}.  The \gamgam\ and
\gampom\ interactions are out of phase, so the interference angle
comes from the real part of the Pomeron and phase shift of the vector
meson in the nuclear potential.

\section{Experimental Feasibility}

Measurements of coherent interactions are only possible if the signals
can be separated from incoherent backgrounds\cite{photon95}.  This
must be possible in both the final analysis and also at the trigger
level; the latter appears to be the harder problem.  The major
backgrounds to be rejected are grazing nuclear collisions,
photo-nuclear interactions, beam gas events, debris from upstream beam
breakup, and cosmic ray muons; the latter two only affect triggering.

STAR (The Solenoidal Tracker at RHIC), is a general purpose large
acceptance detector\cite{STAR}. Time projection chambers track charged
particles with pseudorapidity $|\eta|<2$ and $2.5 < | \eta | < 4$.  A
silicon vertex tracker, time of flight system and TPC $dE/dx$ help
with particle identification.  An electromagnetic calorimeter covers
the range $-1<\eta<2$.  

\begin{figure} 
\centerline{
\epsfig{figure=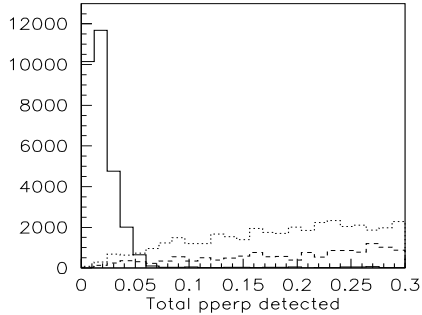,height=1.8 in,width=2.6 in}
}
\caption{ Comparison of $p_\perp$ between \gamgam and FRITIOF
background events passing our cuts. The solid curve is for $\rho^0
\rho^0$ production near threshold, the short dashes are beam-gas and
the long dashes are peripheral nuclear backgrounds.}
\label{fig:rhoplot}
\end{figure}

The coherent event trigger algorithms are based on requiring two or
four tracks in the central TPC, with nothing else visible in the
detector.  Triggering uses a different set of detectors.  The initial
trigger selection uses scintillators and wire chambers (the anode
wires in the TPC endcaps, with a fast readout) surrounding the TPC to
select events based on multiplicity and topology.  Two to five
possible tracks are required, with a reasonable topology.  Timing may
be used to help reject cosmic ray muons and beam gas.  Higher level
trigger algorithms use better hit location information to select
events with two or four charged tracks, with tighter topological cuts.
Then, the calorimeter and TPC tracking contribute, allowing cuts on
neutral multiplicity, total charge, vertex position, and
$p_\perp$. Monte Carlo studies have shown that these trigger
algorithms have good acceptance for coherent events, and adequate
background rejection.

We have simulated the \gamgam\ signals and backgrounds from grazing
nuclear and beam gas\cite{starnote}. We generated tables of \gamgam\
luminosity as a function of invariant mass and rapidity, and then
created events based on these tables.  $p_\perp$ spectra are included
with a $1/R$ Gaussian form factor.  Backgrounds were simulated with
both the FRITIOF and Venus Monte Carlos; the $p_\perp$ spectra are
compared in Fig. 2.

In order to study the effectiveness of different analysis techniques,
we considered three sample analyses that are representative of a wide
range of reactions; these analyses are listed in Table 1.  The
$f_2(1270)$ is a well understood $q\overline q$ meson, representative
of a wide class of particles that decay to two charged particles.  A
spin 0 particle with the same mass and decay channel would have a
slightly lower acceptance.
$\rho^0\rho^0\rightarrow\pi^+\pi^-\pi^+\pi^-$ near threshold (1.5 GeV$
< M_{\rho\rho}<1.6$ GeV) is both a very interesting physics signal,
and also a good benchmark for medium mass processes decaying to four
charged particles.  Finally, the 2960 MeV $J^{PC}=0^{-+}$ $c \overline
c$ resonance $\eta_c$ was chosen as a real challenge; the luminosities
are falling and the branching ratios are small.

For each analysis, we developed a set of cuts based on the required
charged and neutral multiplicity and event topology.  We required
$p_\perp < 100$ MeV, and an appropriate invariant mass and found
the rates and backgrounds given in Table \ref{tab:rates}.  Although
the FRITIOF and Venus background predictions are different, this
analysis shows that $f_2(1270) \rightarrow \pi^+ \pi^-$ and threshold
$\gamma \gamma \rightarrow \rho^0 \rho^0 \rightarrow \pi^+ \pi^- \pi^+
\pi^-$ reactions should be clearly separable from backgrounds, while
the $\eta_c \rightarrow K^{*0}K^-\pi^+$ may be accessible with
particle identification.

\begin{table}
\caption{Rates and backgrounds for \gamgam events for gold on gold
collisions at RHIC for 3 sample analyses.  The $\rho^0\rho^0$ events
were near threshold, with invariant masses between 1.5 and 1.6
GeV/c$^2$. The quantities in parenthesis assume particle 
identification by $dE/dx$ and TOF.}
\begin{tabular}{|l|r|r|r|r|}
\hline 
Channel & Efficiency & Detected & FRITIOF & Venus \\
 & & Events/Yr & Background & Background \\
\hline
$f_2(1270) \rightarrow \pi^+ \pi^-$ & 85\% & 
920,000  & 53,000 & 100,000\\
$\rho^0 \rho^0 \rightarrow  \pi^+ \pi^- \pi^+ \pi^-$ & 38\% &  
16,000 & 3,500 & 1,400 \\
$\eta_c \rightarrow K^{*0}K^-\pi^+$  (w/ PID)& 57\% & 
70 &  210 (8) & 510 (20)\\ 
\hline
\end{tabular}
\label{tab:rates}
\end{table}
We would like to thank our colleagues in the STAR collaboration.  
This work was supported by the DOE under contract DE-AC-03-76SF00098.
E.S. has been partially supported by an NSF fellowship.

\end{document}